\documentclass[twocolumn]{aastex63}

\usepackage{multirow}
\usepackage{rotating}
\usepackage[graphicx]{realboxes}

\newcommand{\lsun}{\mbox{L}_\odot}

\newcommand{\msun}{\mbox{M}_\odot}
\newcommand{\lacc}{L_{\rm acc}}
\newcommand{\macc}{\dot{M}_{\rm acc}}
\newcommand{\lstar}{L_\star}
\newcommand{\mstar}{M_\star}
\newcommand{\rstar}{R_\star}

\definecolor{green}{rgb}{0.0, 0.5, 0.0}

\usepackage{lineno}
\received{}
\revised{}
\accepted{\today}
\submitjournal{AJ}

\shorttitle{WX Cha}
\shortauthors{Fiorellino et al.}
\graphicspath{{./}{figures/}}

\begin{document}

\title{Accretion and extinction variations in the low-mass pre-main sequence binary system WX\,Cha\footnote{Based on observations collected at the European Southern Observatory under ESO programmes 2103.C-5025, 0103.A-9008, and 0100.C-0708(A).}}

\author[0000-0002-5261-6216]{Eleonora Fiorellino}
\email{eleonora.fiorellino@inaf.it}
  \affiliation{Konkoly Observatory, Research Centre for Astronomy and Earth Sciences, E\"otv\"os Lor\'and Research Network (ELKH), Konkoly-Thege Mikl\'os \'ut 15-17, 1121 Budapest, Hungary}
  \affiliation{CSFK, MTA Centre of Excellence, Konkoly-Thege Mikl\'os \'ut 15-17, 1121 Budapest, Hungary}
  \affiliation{INAF-Osservatorio Astronomico di Capodimonte, via Moiariello 16, 80131 Napoli, Italy}
  
\author[0000-0002-4612-5824]{Gabriella Zsidi}
  \affiliation{Konkoly Observatory, Research Centre for Astronomy and Earth Sciences, E\"otv\"os Lor\'and Research Network (ELKH), Konkoly-Thege Mikl\'os \'ut 15-17, 1121 Budapest, Hungary}
  \affiliation{CSFK, MTA Centre of Excellence, Konkoly-Thege Mikl\'os \'ut 15-17, 1121 Budapest, Hungary}
  \affiliation{ELTE E\"otv\"os Lor\'and University, Institute of Physics, P\'azm\'any P\'eter s\'et\'any 1/A, 1117 Budapest, Hungary}

\author[0000-0001-7157-6275]{\'Agnes K\'osp\'al}
  \affiliation{Konkoly Observatory, Research Centre for Astronomy and Earth Sciences, E\"otv\"os Lor\'and Research Network (ELKH), Konkoly-Thege Mikl\'os \'ut 15-17, 1121 Budapest, Hungary}
  \affiliation{CSFK, MTA Centre of Excellence, Konkoly-Thege Mikl\'os \'ut 15-17, 1121 Budapest, Hungary}
  \affiliation{Max Planck Institute for Astronomy, K\"onigstuhl 17, 69117 Heidelberg, Germany}
  \affiliation{ELTE E\"otv\"os Lor\'and University, Institute of Physics, P\'azm\'any P\'eter s\'et\'any 1/A, 1117 Budapest, Hungary}

\author[0000-0001-6015-646X]{P\'eter \'Abrah\'am}
  \affiliation{Konkoly Observatory, Research Centre for Astronomy and Earth Sciences, E\"otv\"os Lor\'and Research Network (ELKH), Konkoly-Thege Mikl\'os \'ut 15-17, 1121 Budapest, Hungary}
   \affiliation{CSFK, MTA Centre of Excellence, Konkoly-Thege Mikl\'os \'ut 15-17, 1121 Budapest, Hungary}
  \affiliation{ELTE E\"otv\"os Lor\'and University, Institute of Physics, P\'azm\'any P\'eter s\'et\'any 1/A, 1117 Budapest, Hungary}
  
\author[0000-0002-8585-4544]{Attila B\'odi}
\affiliation{Konkoly Observatory, Research Centre for Astronomy and Earth Sciences, E\"otv\"os Lor\'and Research Network (ELKH), Konkoly-Thege Mikl\'os \'ut 15-17, 1121 Budapest, Hungary}
\affiliation{CSFK, MTA Centre of Excellence, Konkoly-Thege Mikl\'os \'ut 15-17, 1121 Budapest, Hungary}
\affiliation{MTA CSFK Lend\"ulet Near-Field Cosmology Research Group\\}

\author[0000-0003-3547-3783]{Gaitee Hussain}
  \affiliation{European Space Agency (ESA), European Space Research and Technology Centre (ESTEC), Keplerlaan 1, 2201 AZ Noordwijk,The Netherlands}

\author[0000-0003-3562-262X]{Carlo F. Manara}
  \affiliation{European Southern Observatory, Karl-Schwarzschild-Strasse 2, 85748 Garching bei München, Germany}

\author[0000-0001-5449-2467]{Andr\'as P\'al}
\affiliation{Konkoly Observatory, Research Centre for Astronomy and Earth Sciences, E\"otv\"os Lor\'and Research Network (ELKH), Konkoly-Thege Mikl\'os \'ut 15-17, 1121 Budapest, Hungary}
\affiliation{CSFK, MTA Centre of Excellence, Konkoly-Thege Mikl\'os \'ut 15-17, 1121 Budapest, Hungary}
\affiliation{ELTE E\"otv\"os Lor\'and University, Institute of Physics, P\'azm\'any P\'eter s\'et\'any 1/A, 1117 Budapest, Hungary}

\begin{abstract}

  Light curves of young star systems show photometric variability due to different kinematic, and physical processes.
  One of the main contributors to the photometric variability is the changing mass accretion rate, which regulates the interplay between the forming young star and the protoplanetary disk. 
  We collected high-resolution spectroscopy in eight different epochs, as well as ground-based and space-borne multi-epoch optical and infrared photometry of WX\,Cha, an M0 binary system, with an almost edge-on disk ($i =87^\circ$) in the Chamaeleon\,I star-forming region. 
  Spectroscopic observations cover 72\,days, the ground-based optical monitoring covers 42\,days while space-borne TESS photometry extends for 56\,days. 
  The multi-wavelength light curves exhibit quasi-periodic variability of $0.35-0.53$\,mag in the near-infrared, and of 1.3\,mag in $g$ band. We studied the variability of selected emission lines that trace the accretion, computed the accretion luminosity and the mass accretion rate using empirical relations and obtained values between $\lacc \sim 1.6 \, \lsun \, - \, 3.2 \, \lsun$ and $\macc \sim 3.31 \times 10^{-7} \,\msun/{\rm yr} \, - \, 7.76 \times 10^{-7}$\,$\msun/$yr.  
  Our results show that WX\,Cha is accreting at a rate larger than what is typical for T\,Tauri stars in the same star-forming region with the same stellar parameters. We theorize that this is due to the higher disk mass of WX\,Cha than what is usual for stars with similar stellar mass, and to the binary nature of the system.
  Daily changes in the accretion luminosity and in the extinction can explain the photometric variability. 

\end{abstract}


\section{Introduction} \label{sect:intro}

According to the current star formation scenario, the material flows from a dusty infalling envelope and the circumstellar disk to the star. 
The magnetospheric accretion model describes the accretion process from the inner disk to the stellar surface on single stars \citep[e.g][]{har16}. 
This mechanism drives the evolution of the disk, the star, and the planet formation \citep[e.g.,][]{manaraPPVII}. 
For binary systems this scenario is complicated by the interplay of the stellar components, their circumstellar disks, and the possible presence of a circumbinary disk (CBD) \citep[e.g.,][]{offnerPPVII}. 
Young stellar objects (YSOs) show a certain degree of variability which is typically related to the accretion process \citep[e.g.,][]{fischerPPVII}. 
However, YSOs variability can also be due to other stellar events, such as circumstellar obscuration, hot or cold spots on the star and/or disk, ejection of jets, stellar winds, and
rapid structural changes in the inner disk \citep[][]{cody2014}.
To understand the variability nature of young stars, and, in particular, how much the accretion process affects the variability of binaries, each system has to be studied in detail with high resolution spectroscopy and contemporary photometry, for many epochs. 
The most suitable candidates for such study are located in the nearby ($< 500$\,pc) star-forming regions.

WX\,Cha (2MASS J11095873-7737088) is a young binary system, located at $d=189.1^{+1.9}_{-2.2}$\,pc \citep{bai21} in the Chamaeleon\,I (Cha\,I) star-forming region. 
This system was recognized as a binary for the first time by
\citet[][]{vog12}. 
They showed that its components form a physical pair with a projected separation of 0.75\,arcsec (141\,au with the new Gaia distance) whose flux ratios are 8.1 in the $H$-band and 10.2 in the $K$-band. 
They remarked also about the $K$-band photometric variability of the system.
Concerning the evolutionary stage of WX\,Cha, \citet[][]{luh08} compiled its optical-infrared spectral energy distribution (SED) and classified it as a Class\,II YSO, or Classical T\,Tauri star (CTTS), based on the spectral index of $\alpha = -1.10$ between 3.6\,$\mu$m and 24\,$\mu$m.
Later, \citet[][]{dae13} resolved the system using near-infrared (NIR) adaptive optics images obtained with VLT/NACO. 
Based on this, they computed flux ratios between the primary and the secondary component of 5.1, 6.7, 9.4, and 16.4 in the $J$, $H$, $K$, $L'$ bands, respectively. 
They classified the primary component as an M1 star with $T_{\rm eff}=3710$\,K, $\mstar=0.49$\,$\msun$, $A_V = 3.2 \pm 0.2$\,mag, and $\macc = (98 \pm 35) \times 10^{-9}$\,$\msun$yr$^{-1}$; and the secondary component as an M5 star with $T_{\rm eff}=3130$\,K, $\mstar=0.18$\,$\msun$, $A_V = 2.3 \pm 0.2$\,mag, and $\macc < 0.58 \times 10^{-9}$\,$\msun$yr$^{-1}$, assuming a distance of $160 \pm 15$\,pc. 
\cite{banzatti2015} determined an inner disk inclination of 87$^\circ$ using CRIRES measurements.
\citet[][]{dae13} also computed the probability of each component to have an accretion disk, calculated
for each target component from the measured Br$\gamma$ equivalent width, its uncertainty, and the local noise. They reported in their Tab.\,4 that this probability for the primary is 1.00 (certain), and the probability for the secondary component is 0.01 (very unlikely). Therefore, the secondary component probably does not have a disk and is not accreting. 
ALMA archival data show that the disk of the primary was detected but not spatially resolved and the disk of the secondary not resolved. However there are deeper, higher resolution ALMA observations indicated as {\it in progress} in the archive. These will hopefully reveal more about the disk of the primary in the future. 
The observations available at the moment suggest that if there was ever an interaction between the two disks, it must have occurred in the past, and that the secondary does not have an accreting circumstellar disk.
\citet[][]{pascucci16} provided ALMA flux density $F_\nu = (20.81 \pm 0.57)$\,mJy at 338\,GHz (887\,$\mu$m), which corresponds to $M_{dust} = 7.78$\,$M_\Earth$ using the updated distance \citep[][]{manaraPPVII}.
The most recent estimates of the stellar and accretion parameters are $\mstar = 0.491$\,$\msun$, $\lstar = 0.86 $\,$\lsun$, accretion luminosity of $\log (\lacc/\lsun) = 0.0341$, and mass accretion rate of $\log (\macc /\msun$yr$^{-1}) = -6.729$ \citep[][]{manaraPPVII}, for a M0.5 spectral type, $A_V = 3.0$\,mag \citep[][]{man17cha}. 
These results are based on X-Shooter spectroscopy, where the system was not resolved, although these values are probably closer to the values of the primary, as that dominates the spectrum of the system. 

Although WX\,Cha is often among the targets of surveys aimed at the Cha\,I star-forming region \citep[e.g.][]{har98, cutri2003, luh07, luh08, ngu12, vog12, costigan12, dae13, frasca15, man17cha, man19}, no recent study is dedicated to its detailed analysis, in particular to its light curve variability.
In this work, we fill this gap by investigating for the first time the photometric and spectroscopic variability behavior of this binary system and comparing our results with other multiple systems and single young stars in the same region. 
For this purpose, we used multi-epoch optical and NIR photometry and high-resolution optical spectroscopy, and analyzed the relation between the photometric variability and changes in the spectral lines. 

The paper is structured as follows. In Sect.\,\ref{sect:obs&data} we describe the observations and data reduction; in Sect.\,\ref{sect:results} we present main results, such as the system parameters, line flux measurements, and accretion rates; in Sect.\,\ref{sect:discussion} we discuss our results. We summarize our conclusions in Sect.\,\ref{sect:conclusions}.


\begin{figure*}[t]
    \centering
    \includegraphics[width=\textwidth]{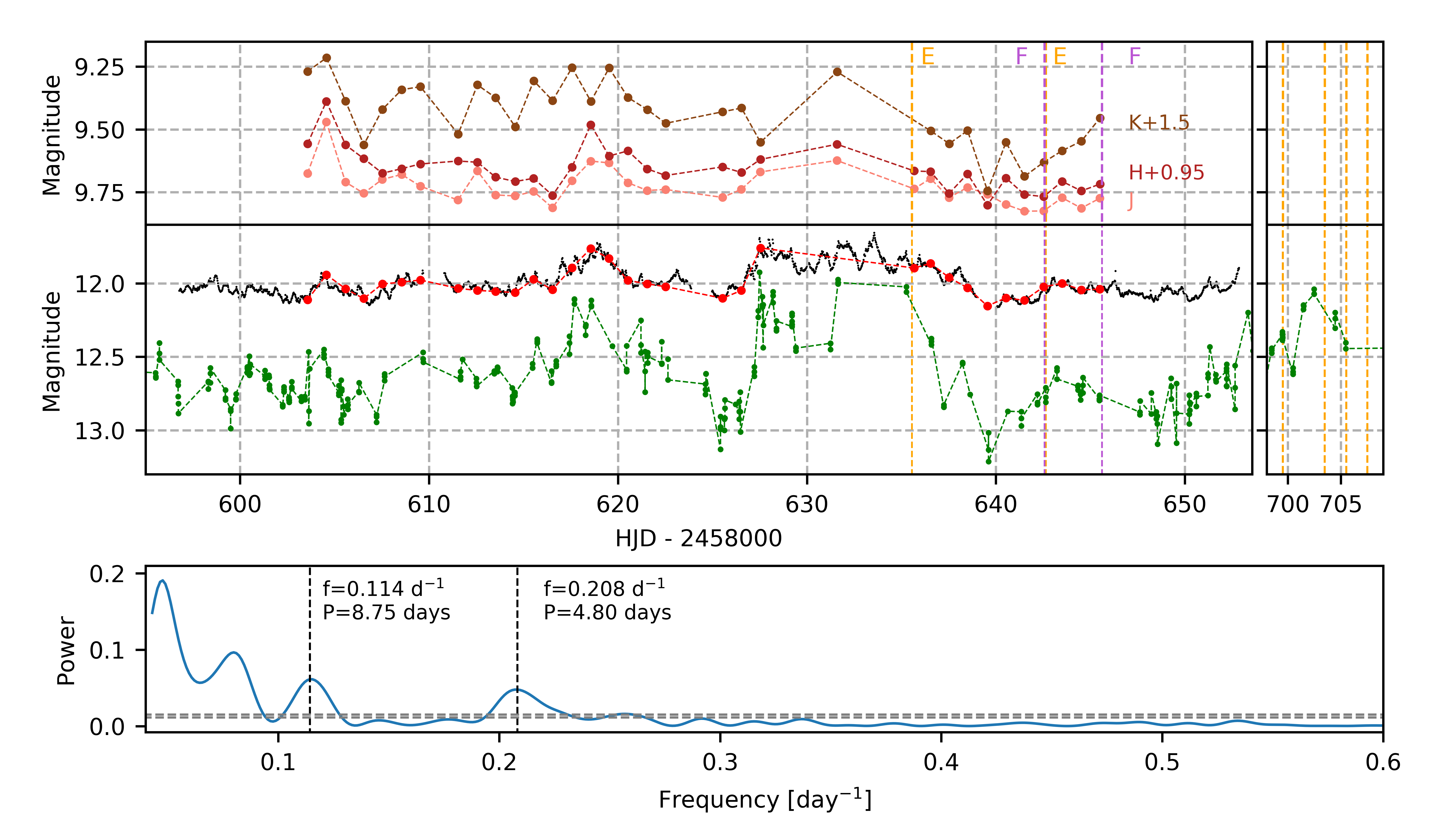}
    \caption{The 2019 (top panel) TESS light curve is indicated with black dots, and the contemporaneous ASAS\,SN $g$\,band light curve is shown with green points. SMARTS $I$-band (red), $J$ (pink), $H$ (bordeaux), and $K$ (brown) light curves are also added.
    Vertical dashed lines correspond to epochs when ESPRESSO (orange) and FEROS (magenta) spectra were taken. Since Epoch 2 (FEROS) and 3 (ESPRESSO) were taken during the same night, so these epochs are superimposed in the figure.
    The blue line (bottom panel) shows the Lomb-Scargle periodogram obtained using the 2019 TESS data. Points of the same light curve are linked with dashed lines only to help the reader follow each light curve. Horizontal lines represent false alarm probabilities of 1\%, and 0.01\%, respectively.}
    \label{fig:lc_period}
\end{figure*}

\begin{figure*}[t]
    \centering
    \includegraphics[width=\textwidth]{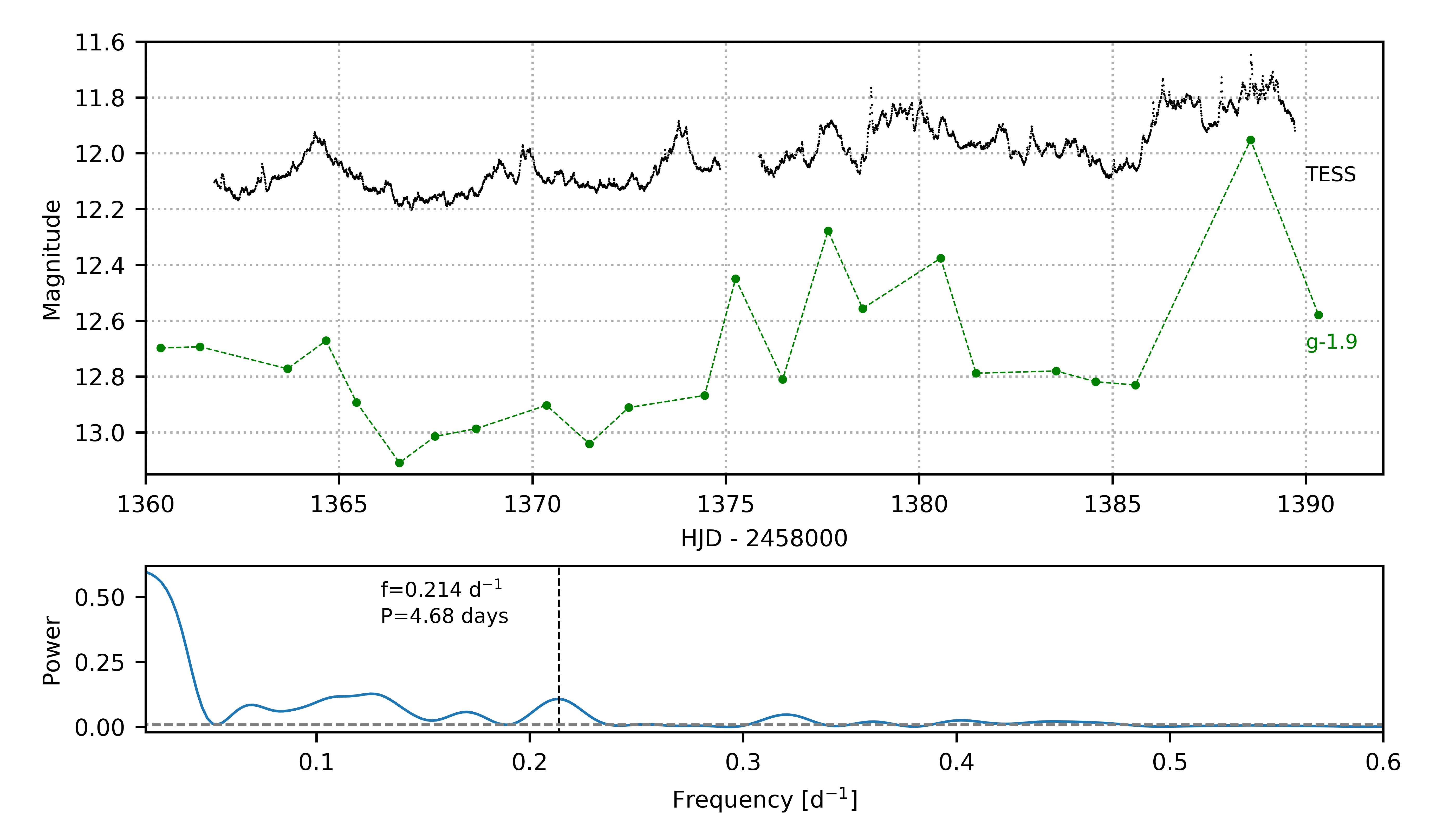}
    \caption{The 2021 (top panel) TESS light curve is indicated with black dots, and the contemporaneous ASAS\,SN $g$\,band light curve is shown with green points. 
    The blue line (bottom panel) shows the Lomb-Scargle periodogram obtained using the 2021 TESS data. Points of the same light curve are linked with dashed lines only to help the reader follow each light curve. Horizontal lines represent false alarm probabilities of 1\%, and 0.01\%, respectively.}
    \label{fig:lc_period_2021}
\end{figure*}

\section{Observations and Data Reduction} \label{sect:obs&data}

This work dedicated to WX\,Cha is part of a large observational effort whose aim was to take full advantage of the Transiting  Exoplanet Survey Telescope \citep[TESS,][]{ricker2015} monitoring of the Cha\,I region in 2019 to study the variability of six selected CTTS.  
For this purpose we organized ground-based multi-wavelength photometric and spectroscopic monitoring contemporaneously with TESS. 
We used the SMARTS 1.3\,m telescope (Cerro Tololo, Chile), VLT/ESPRESSO (Paranal Observatory, Chile), and the MPG2.2m/FEROS (La Silla Observatory, Chile). 
The data reduction has been performed in the same way for the whole sample, and it is discussed in detail in \citet[][]{zsidi2022}. 
Here, we briefly summarize the main steps.

\subsection{Photometry}
TESS provided a 56-day-long light curve of WX\,Cha with 30\,minutes cadence between 2019 April 23 and 2019 June 18, and a 28-day-long light curve with 10\,minutes cadence between 2021 May 27 and 2021 June 24.   
To extract the photometry, we used an aperture radius of 2\,pixels (40$\arcsec$), and the average $I_C$ magnitude of WX\,Cha from our SMARTS monitoring as the reference brightness. 

We obtained ground-based photometric monitoring between 2019 April 30 and 2019 June 11. We used the SMARTS/ANDICAM optical-infrared imager mounted on the 1.3\,m telescope at Cerro Tololo (Chile), and took $I_CJHK$ images with an approximately nightly cadence. 
We also took $VR_C$ images for the first three nights. 
In the optical, we typically obtained 21 images with 14\,s exposure time and calculated aperture photometry for WX\,Cha and a comparison star, UCAC3\,25-23448, using an aperture radius of six pixels ($2.23\arcsec$) and sky anulus between 10 and 15 pixels (3.71$\arcsec - 5.57 \arcsec$).
Photometric calibration was done using the APASS9 magnitudes \citep[][]{hen16} of the comparison star that we converted to $R_C I_C$ using the conversion formulae of \citet[][]{jor06}.
In the infrared (IR) we obtained $5-9$ ($J-$ and $H-$bands) or $5-17$ ($K-$band) images, with exposure times between 4\,s and 60\,s. 
We performed dithering to remove bad pixels and the sky signal, which we perfomed using our custom procedures written in IDL. 
Reduced flat-field images were provided by the SMARTS consortium. 
We perfomed aperture photometry with an aperture radius of 6 pixels (1.6$\arcsec$) and sky radius between 18 and 24 pixels (4.8$\arcsec$ and 6.5$\arcsec$), except for WX\,Cha where a larger sky area was used due to contamination by a nearby star. 
For the photometric calibration in the infrared, we used 2MASS magnitudes \citep{cutri2003} of four comparison stars, although not all four were visible on all frames due to a repositioning of the $2\farcm4\times2\farcm4$ field of view during the observing campaign.
The resultant light curves are shown in Figs.\,\ref{fig:lc_period} and \ref{fig:lc_period_2021}. 
See 
Appendix\,\ref{app:smarts} for further details. 

\subsection{Spectroscopy} \label{sect:spec}
We carried out high-resolution ($R=140\,000$) optical $(380-788\,$nm) spectroscopic observations with ESPRESSO as part of the DDT proposal (Pr.Id.2103.C-5025, PI \'A. K\'osp\'al) between 2019 June and August, see Tab.\,\ref{tab:log_obs}. 
We carried out the bias, dark, and flat-field corrections and the wavelength calibration using Version 3.13.2 of the EsoReflex/ESPRESSO pipeline \citep{fre13}. 
The resulting 2D spectra were merged producing a rebinned 1D spectra.
We used the default parameters for each step of the pipeline. 

We obtained two additional high-resolution ($R=48\,000$) optical ($350-920\,$nm) spectra on June 8 and June 11 with FEROS in object-calib mode, where contemporaneous spectra of a ThAr lamp were recorded throughout the whole object exposure. 
The \textsc{ferospipe} pipeline in \textsc{python} is available for reducing data acquired with the FEROS instrument. 
A detailed description of the pipeline can be found in \cite{bra17}. 
However, this pipeline was designed to precisely measure the radial velocity, and it calibrates only 25 of the available 33 \'echelle orders. 
The excluded orders are essential for our analysis as they cover the $\sim 673 - 823$\,nm range, which includes multiple accretion tracer lines. 
Therefore, we modified the original pipeline in a way that we extended the wavelength coverage to all orders. 

The flux calibration of all spectra was performed in two steps. First, the spectra were normalized. 
Then, the calibration coefficients were calculated using the available optical photometry.
We interpolated the optical brightness of WX\,Cha for the exact epochs of the spectroscopic observations in the ASAS-SN $g-$band ($\lambda_{\rm eff} = 470$\,nm) and SMARTS I$_C-$band ($\lambda_{\rm eff} = 784$\,nm) light curves, providing calibration coefficients pixel by pixel. 
We checked the reliability of the coefficients by applying them to the $g$-band photometry. We found a relative error on the $I$-band magnitude of 0.003 on average.
This was possible for the first four epochs, which were covered by our SMARTS monitoring. 
For the subsequent epochs, we interpolated the available photometry in $g$-band, using the results as an estimate of the contemporary photometry.
We obtained m$_g = 14.46$\,mag, 14.16\,mag, 14.42\,mag for epochs\,5, 6 and 7, respectively. 
The ASAS-SN observations for WX\,Cha have been interrupted for several months, so it was not possible to determine the photometry of our target in the last epoch (Epoch\,8) by interpolating, not even for the $g$-band. 
For this reason, to flux calibrate the spectrum of Epoch\,8, we used a period of $P = 9.6$\,days (see Sect.\,\ref{sect:period}) to extrapolate the available photometry for this epoch, obtaining m$_g = 14.70$\,mag.
Magnitudes were converted into fluxes by using zero magnitude fluxes\footnote{\url{http://svo2.cab.inta-csic.es/theory/fps/}}. 
The calibrated spectra were obtained by multiplying the normalized spectra by the calibration function. 
The flux calibrated spectra are shown in 
Appendix\,\ref{app:spectra}.

\begin{deluxetable}{llcc}[t]
\tablecaption{\label{tab:log_obs}}
\tablewidth{0pt}
\tablehead{
\colhead{Epoch} & Date & HJD        & Instrument \\
\colhead{}     &     & (+2458000) & }
\startdata
Epoch\,1 & 2019 Jun 01  & 635.549  & ESPRESSO\\
Epoch\,2 & 2019 Jun 08  & 642.571 & FEROS \\
Epoch\,3 & 2019 Jun 08  & 642.639  & ESPRESSO\\
Epoch\,4 & 2019 Jun 11 & 645.604 & FEROS \\
Epoch\,5 & 2019 Aug 04  & 699.530 & ESPRESSO\\
Epoch\,6 & 2019 Aug 07  & 703.481 & ESPRESSO\\
Epoch\,7 & 2019 Aug 10 & 705.524 & ESPRESSO\\
Epoch\,8 & 2019 Aug 11 & 707.498 & ESPRESSO\\
\enddata
\end{deluxetable}

\section{results}
\label{sect:results}

\subsection{The Binary Components as Seen by Gaia} \label{sect:gaia}

In order to investigate which component contributes to the observed flux and the variability, we examined the Gaia\,EDR3 measurements \citep[][]{gaia_citation1, gaia_citation2, gaia3}. 
Gaia\,EDR3\,5201148946703195904 and 5201148946701210624 correspond to WX\,Cha A and B, respectively.
According to these data, the separation of the two components is 0.7425
$\arcsec \pm 0.0014 \arcsec$, their P.A. is $52.764^\circ \pm 0.012^\circ$. 
Both results are in agreement with parameters derived earlier by \citet[][]{dae13}: $ a= 0.74 \pm 0.01 \arcsec$, $\mbox{P.A.}= 51.8^\circ \pm0.5^\circ$.

Gaia G magnitudes are $m_G=13.23$\,mag and 15.94\,mag for WX\,Cha A and B, respectively. Thus, the primary is about 12 times brighter on average than the companion, i.e. 92\% 
of the total flux comes from the primary and 8\% 
of the total flux comes from the companion. There is no proper motion for the companion (nor BP, nor RP magnitudes, nor parallax).
The Gaia PSF is about 0.177$\arcsec$, thus the WX\,Cha components are well resolved by Gaia.

\subsection{Light Curves} \label{sect:period}
Fig.\,\ref{fig:lc_period} shows the multi-filter light curves of WX\,Cha for the year 2019. 
Irregular variations on a daily scale, with peak-to-peak amplitude of 0.5\,mag are present in the TESS light curve (black points). 
The irregular behavior of the light curve is also revealed by the two maxima, which have different shapes: the first one (HJD = 2458618) consists in a single peak of few days, while the second one corresponds to a plateau in the brightness of about two weeks (HJD = $2458627 - 2458635$), which ends with the dimmest epoch.
The peaks and minima of the TESS light curve are present also in ASAS-SN $g-$band and in the SMARTS ($I_CJHK$ bands) light curves for year 2019. 
The $g$ band light curve presents a third peak around HJD = 2458703. 
We note that the peak-to-peak amplitude varies with the wavelength, being more pronounced in $g$ band (1.3\,mag) and smaller in NIR ($0.35$\,mag in $J$  and $H$ bands, and 0.53\,mag in $K$ band).
Light curves for the year 2021 also present irregular behavior with a peak-to-peak amplitude similar to that observed in 2019 of 0.55\,mag in the TESS light curve and
of 1.2\,mag in the $g$ band light curve. 
Fig.\,\ref{fig:lc_period_2021} shows two peaks in $g$ band, the first which looks like a series of bursts (HJD = $2459378 - 2459383$), and the second resembling a plateau (HJD = $2459386 - 2459389$). 
The corresponding TESS light curve seems to confirm the series of burst during the first peak, shows a plateau in the brightness during the second, and suggests the presence of another peak at HJD = 2459363.

We computed a Lomb-Scargle periodogram \citep[][]{lom76, sca82} to look for any periodic signals. 
The bottom panels of Figs.\,\ref{fig:lc_period} and \ref{fig:lc_period_2021} show such periodograms obtained from the 2019 and 2021 TESS light curves, respectively. 
For the 2019 TESS data, we calculated the levels corresponding to the 1\% and the 0.01\% false alarm probabilities (FAP), and we show the results with horizontal dashed lines in the bottom panel of Fig.\,\ref{fig:lc_period}. This shows that there are two long-term trends under 0.1\,day$^{-1}$. We, however, do not discuss these trends here, as they are longer than the expected rotational periods for classical T Tauri stars. We detected two additional peaks at 8.75 days and 4.80\,days, which fall in the range of typical rotational periods.
We carried out a similar period analysis on the ASAS-SN data as well. 
These data show several peaks, two of them are close to the values presented by the TESS data: one peak is at 4.8\,days, and the other is at 9.6\,days. 
Since the latter period seems to better reproduce the $g$ band light curve, see for example the first peak at HJD\,=\,2458618 and the second one after 9 days, we used this period to extrapolate the $g$ band magnitude for Epoch\,8 (see Sect.\,\ref{sect:spec}).
We followed the same procedure for the 2021 TESS light curve as well. This resulted in a peak at 4.68\,day, while the presence of the longer period around 9\,day is not confirmed as the peak is very smeared out.
We also performed consecutive pre-whiting steps, then after pre-whitening with the highest amplitude peak (the 9.6\,day period), we found three significant peaks: the first two peaks correspond to 12.4\,days and 8.1\,day, respectively, and the third one is the 4.6\,day.
We note that the 9.6\,day period is twice the $4.6-4.8$\,day period. 
It is possible that either the former or the latter is the rotational period of the primary star. 
However, we point out that, despite this analysis, the WX\,Cha system do not show regular variability according to our data set and that our choice is driven by the will to flux calibrate the last spectroscopic Epoch. Indeed, we alert the reader about the uncertainty that this assumption propagates on the last epoch results. 

According to \citet[][]{cody2014}, a light curve can be described by two parameters. The $Q$ parameter provides information about the periodicity of the light curve (if $Q=0$ the light curve is perfectly periodic, while, if $Q = 1$ the light curve is stochastic), and the $M$ parameter provides information about the symmetry of the light curve ($M = -1$ if the light curve is asymmetric and dominated by burst, $M \sim 0$ if it is symmetric, and $M = 1$ if it is asymmetric and dominated by dippers). 
By applying the same method described in \citet[][]{cody2014} to TESS light curves, we found that coefficients are $Q = 0.77$ and $M=-0.19$ for the 2019 light curve, and $Q=0.60$ and $M = -0.43$ for the 2021 light curve. Using the thresholds shown in Fig.\,31 of \citet[][]{cody2014}, we interpreted our results as follows: the light curve is always aperiodic, in agreement with what we found from the LS periodogram analysis, i.e. it is not possible to determine the period of WX\,Cha; the light curve is symmetric during 2019, even if the $M$ value is close to the bursting threshold, and becomes asymmetric dominated by bursts in 2021.

\subsection{Extinction and Spectral Typing} \label{sect:extiction}
\begin{figure}[t]
    \centering
    \includegraphics[width=\columnwidth]{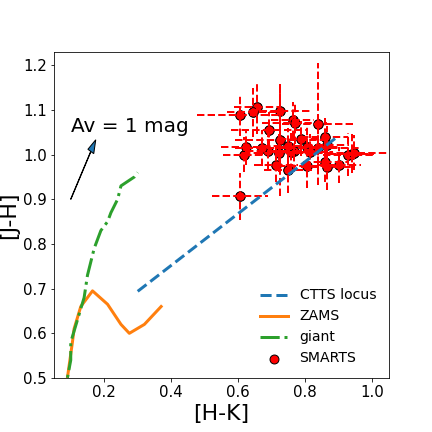}
    \caption{Color-Color diagram of WX\,Cha. Red circles are data from SMARTS observations at different epochs. Blue dashed line represent the CTTS locus. The black arrow shows the shift from the CTTS locus for a source with 1\,mag of extinction. The orange solid line
and the green dot-dashed line correspond to the ZAMS and
the giant branch, respectively.}
    \label{fig:colcol_diag}
\end{figure}
The interstellar extinction is one crucial parameter in our analysis, necessary to deredden the line fluxes and to calculate the accretion rate. 
The variation of the extinction can also play a role in the photometric variability.

Several $A_V$ estimates are available in the literature for WX\,Cha. 
\citet[][]{luh07} estimated $A_J = 0.56$\,mag, for the primary source, by converting the color excess between 0.6\,$\mu$m and 0.9\,$\mu$m. 
This corresponds to $A_V = 1.97$\,mag, using the extinction law by \citet[][]{car89} using $R_V = 3.1$. 
More recently, \citet[][]{dae13} estimated $A^1_V = 3.2 \pm 0.2$\,mag and $A^2_V = 2.3 \pm 0.2$\,mag for the primary and secondary components, respectively. 
They provided these results using color-color diagrams, as it is described below in this section. 
Lastly, \citet[][]{man17cha} obtained $A_V = 3.0 \pm 0.3$\,mag and M0.5 spectral type, by using an automatic routine which analyze the spectra, comparing it with a non accreting template and a slab model, providing the spectral type, the extinction value, and stellar parameters.

We computed $A_V$ by projecting the position of WX\,Cha back to the CTTS locus \citep{mey97} in the [J$-$H]\,vs.\,[H$-$K] color-color diagram following the extinction-vector by \citet{car89} and $R_V=3.1$ (Fig.\,\ref{fig:colcol_diag}). 
The mean value of the overall photometry displayed in the color-color diagram corresponds to $A_V = 0.71 \pm 0.30$\,mag, which is not in agreement with any of the previous extinction estimates. This is probably due to the large uncertainties on the K-band photometry. Moreover, the disagreement between the extinction obtained by the spectral typing using the visible spectra, and by using the color-color diagrams relation can be due to the fact that the H and K-band magnitudes are affected by disk emission.

We also computed the extinction after having performed the spectral typing of WX\,Cha. 
We compared the absorption lines, setting a threshold to 80\% of the flux, of a set of main sequence stellar templates from \citet[][]{coe05}, from $T_{\rm eff} = 3500$\,K to $7000$\,K to our observed spectrum at Epoch\,1, masking the emission lines, in the wavelength range $600 - 800$\,nm. 
We found that the template with $T_{\rm eff} = 3740$\,K minimizes the $\chi^2$ between the synthetic spectrum and the measured spectrum absorption lines.  
This effective temperature corresponds to M0 spectral type according to \citet[][]{pec13}.
Once we determined the spectral type, we fitted $I_C$, $J$, and $H$ data points of the SED with a \citet[][]{pec13} model, by fixing the temperature to $T_{\rm eff} = 3750$\,K. 
We obtained different values of the extinction for each epoch observed with SMARTS data. In the following analysis we adopt their mean value of $A_V = 3.4 \pm 0.2$\,mag. 
In Tab.\,\ref{tab:Av} we report the exact values we estimated for the corresponding spectra, when available.

The $A_V$ is degenerate with the spectral type. 
The latter method to compute the extinction we described takes into account also the spectral type of the source, thus we chose to adopt the value provided with this method in the following. 
Moreover, the result we obtain in this way is in agreement with the literature. 
Indeed, we find that the spectral type we derive, M0, is in agreement within the uncertainty with the M0.5 type obtained by \citet[][]{man17cha}, where UVB, VIS, and NIR data were used, while it is earlier than M1.25 and M1$-$M5 types, estimated by \citet[][]{luh07} and \citet[][]{dae13}, respectively, where only NIR data were analyzed. 
This disagreement is not surprising because the spectral type can be earlier when investigated at shorter wavelengths \citep[e.g.][]{gullysantiago17}. 
The extinction we find $A_V = 3.4 \pm 0.2$\,mag is in agreement within the error with the values provided  by \citet[][$A_V = 3.0 \pm 0.5$]{man17cha} and \citet[][$A_V = 3.2 \pm 0.2$\,mag]{dae13}.

\begin{figure}
    \centering
    \includegraphics[angle=90,width=\columnwidth]{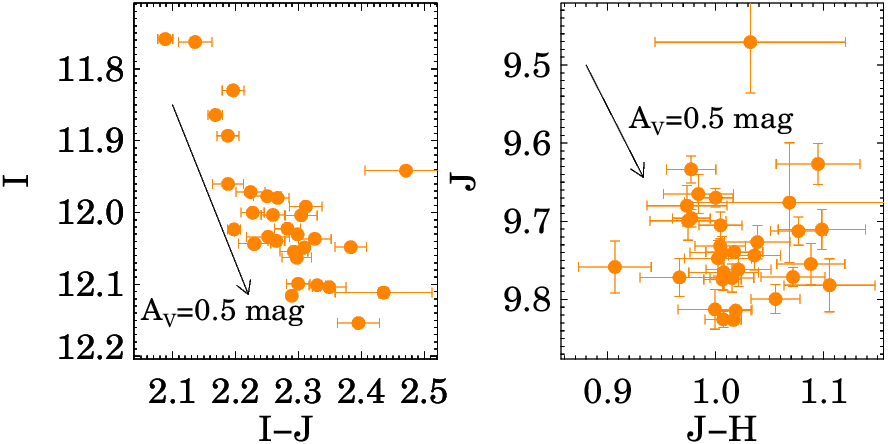}
    \caption{Color-magnitude diagrams of WX\,Cha.}
    \label{fig:cmd}
\end{figure}

\subsection{Color-magnitude Diagrams} \label{sect:col_diag}
To investigate the IR photometric properties of WX\,Cha system, we built  color-magnitude diagrams in Fig.\,\ref{fig:cmd}. 
The left panel shows linear decreasing trend in the [$I$]\,vs.\,[$I-J$] diagram. 
This suggests that the $I$ and $J$ band variations are indeed correlated, but $I$ band variations are larger than $J$ band variations (see light curves on Fig.\,\ref{fig:lc_period}), so the source is redder when fainter, hence the downward trend in the color-magnitude diagram.
This trend looks to be consistent with extinction changes.
The right panel shows a scattered distribution of 0.2\,mag variation on both  axes on the [$J$]\,vs.\,[$J-H$] diagram, which is still not inconsistent with reddening changes.
Unfortunately, our ground-based $K$ band suffers of large uncertainties that prevent us to find any trend, and, therefore, interpret physically the NIR diagrams of Fig.\,\ref{fig:colcol_diag} and \ref{fig:cmd}. 

\subsection{Veiling}
\label{sect:veiling}
The veiling of a star can be calculated by measuring the equivalent width ($W_{\rm eq}$) of absorption lines in the spectra:
\begin{equation}
    W^0_{\rm eq} = W_{\rm eq}(V+1)
\end{equation}
where $W_{\rm eq}$ is the measured equivalent width, $ W^0_{\rm eq}$ is the equivalent width of the line in the unveiled reference spectrum, and $V$ is the veiling. 
We choose to take an M0 template as reference \citep[the ESPaDOnS spectrum of a non accreting Class III YSO identified as M0 PMS star by][ available from the Polarbase database\footnote{\url{http://polarbase.irap.omp.eu/} }]{man13templ}, which is the spectral type as we derived for WX\,Cha (see Sect.\,\ref{sect:extiction}). 
This enabled us to provide the absolute veiling of the system. 
We calculated the veiling of WX\,Cha by using the equivalent widths of the absorption lines in the red part of the spectra ($\lambda = 584-788$\,nm). 
We found values comprised between 0.64 and 1.79, see Tab.\,\ref{tab:Av} and Fig.\,\ref{fig:veiling} (top panel).

\subsection{Accretion Line Fluxes} \label{sect:lineflux}
In the magnetospheric accretion scenario, the accretion of material from the disk onto the central star in a young stellar object is traced by many emission lines in the optical spectra, for example H{\footnotesize{I}}, He{\footnotesize{I}}, Ca{\footnotesize{II}}, and Na{\footnotesize{I}}  (Fig.\,\ref{fig:var_profile}), see \citet[][]{har16} for a review. 
In the last two decades, empirical relations between the luminosity of these emission lines and the accretion luminosity have been developed \citep[e.g.][]{muz98, ant14, alc14}. We used the most recent relations of \citet[][]{alc17} to estimate the accretion rates of WX\,Cha, starting from the line flux measurements of the different accretion tracers. 
Therefore, we searched for lines for which the line luminosity vs. accretion luminosity relation has been 
\begin{rotatetable*}
\startlongtable
\begin{deluxetable*}{lrccccrccc}
\centerwidetable
\tablecaption{\label{tab:flux} Observed flux of accretion tracers before correcting for the extinction.}
 \tablewidth{\textwidth}
\tablehead{
\colhead{element} & $\lambda_{line}$ & $F^{ep1}_{obs}$  & $F^{ep2}_{obs}$(F) & $F^{ep3}_{obs}(E)$ & $F^{ep4}_{obs}$ & $F^{ep5}_{obs}$ & $F^{ep6}_{obs}$ & $F^{ep7}_{obs}$ & $F^{ep8}_{obs}$ \\
\colhead{}        & [nm] &  \multicolumn{8}{c}{[$10^{-14}$ erg s$^{-1}$ cm$^{-2}$]}                }
\decimalcolnumbers
\startdata
H3  &     656.5 &  $ 213.760 \pm  0.096$ & $ 164.263 \pm  0.212$ & $ 185.911 \pm  0.078$ & $ 103.881 \pm  0.439$ & $ 159.347 \pm  0.083$ & $ 123.482 \pm  0.082$ & $ 166.180 \pm  0.182$ & $ 224.002 \pm  0.083$ \\
H4  &     486.3 & $  28.694 \pm  0.121$ & $  14.931 \pm  0.122$ & $  21.350 \pm  0.034$ & $   3.345 \pm  0.144$ & $  25.257 \pm  0.103$ & $  19.523 \pm  0.081$ & $  25.274 \pm  0.300$ & $  31.545 \pm  0.080$ \\
H5  &     434.2 & $   9.843 \pm  0.090$ & $   1.997 \pm  0.107$ & $   6.674 \pm  0.077$ & $   0.676 \pm  0.185$ & $   9.522 \pm  0.077$ & $   8.617 \pm  0.063$ & $  10.973 \pm  0.262$ & $  10.311 \pm  0.050$ \\
H6  &     410.3 & $   7.310 \pm  0.254$ & $   6.495 \pm  0.467$ & $   3.888 \pm  0.043$ & $   9.021 \pm  1.018$ & $   7.037 \pm  0.108$ & $   6.671 \pm  0.083$ & $   7.408 \pm  0.415$ & $   5.927 \pm  0.058$ \\
H7  &     397.1 & $  13.166 \pm  0.200$ & $   7.081 \pm  0.700$ & $   4.756 \pm  0.057$ & $   4.964 \pm  1.132$ & $   7.899 \pm  0.160$ & $   7.759 \pm  0.133$ & $   7.044 \pm  0.556$ & $   8.700 \pm  0.092$ \\
H8  &     389.0 & $   3.088 \pm  0.262$ & $   < 2.317         $ & $   1.180 \pm  0.057$ & $   < 2.963         $ & $   3.514 \pm  0.216$ & $   2.803 \pm  0.174$ & $   3.225 \pm  0.816$ & $   1.996 \pm  0.080$ \\
HeI  &    447.3 &  $   0.951 \pm  0.077$ & $   0.577 \pm  0.100$ & $   0.803 \pm  0.034$ & $   < 1.385         $ & $   1.315 \pm  0.069$ & $   1.068 \pm  0.048$ & $   1.518 \pm  0.231$ & $   0.958 \pm  0.044$ \\
HeI  &    471.5 & $   0.198 \pm  0.024$ & $   < 0.305         $ & $   0.104 \pm  0.008$ & $   0.867 \pm  0.274$ & $   0.134 \pm  0.012$ & $   0.164 \pm  0.018$ & $  < 0.264 $ & $   0.151 \pm  0.016$ \\
HeIFeI  & 492.4 & $   2.152 \pm  0.081$ & $   1.902 \pm  0.079$ & $   1.442 \pm  0.045$ & $   < 0.512         $ & $   2.021 \pm  0.071$ & $   1.175 \pm  0.029$ & $   1.506 \pm  0.124$ & $   2.586 \pm  0.029$ \\
HeI  &    501.7 & $   2.836 \pm  0.101$ & $   2.278 \pm  0.099$ & $   1.930 \pm  0.056$ & $   1.652 \pm  0.279$ & $   2.638 \pm  0.086$ & $   1.295 \pm  0.072$ & $   2.255 \pm  0.314$ & $   3.553 \pm  0.067$ \\
HeI  &    587.7 & $   2.512 \pm  0.037$ & $   2.872 \pm  0.065$ & $   2.907 \pm  0.044$ & $   1.825 \pm  0.113$ & $   3.621 \pm  0.035$ & $   3.139 \pm  0.028$ & $   3.659 \pm  0.089$ & $   3.675 \pm  0.054$ \\
HeI  &    668.0 & $   1.264 \pm  0.023$ & $   1.339 \pm  0.057$ & $   1.428 \pm  0.018$ & $   1.239 \pm  0.109$ & $   1.973 \pm  0.022$ & $   1.770 \pm  0.018$ & $   1.839 \pm  0.056$ & $   1.835 \pm  0.023$ \\
HeI  &    706.7 & $   1.119 \pm  0.024$ & $   4.540 \pm  0.088$ & $   1.229 \pm  0.019$ & $  12.414 \pm  0.190$ & $   1.488 \pm  0.021$ & $   1.358 \pm  0.020$ & $   1.396 \pm  0.054$ & $   1.428 \pm  0.022$ \\
HeI  &    728.4 & $   0.143 \pm  0.015$ & $   0.156 \pm  0.024$ & $   0.187 \pm  0.011$ & $   0.219 \pm  0.052$ & $   0.220 \pm  0.020$ & $   0.192 \pm  0.014$ & $   0.230 \pm  0.026$ & $   0.146 \pm  0.014$ \\
HeII  &   468.7 & $   0.645 \pm  0.058$ & $   0.352 \pm  0.073$ & $   0.354 \pm  0.014$ & $   1.223 \pm  0.291$ & $   0.485 \pm  0.048$ & $   0.579 \pm  0.039$ & $   0.552 \pm  0.170$ & $   0.382 \pm  0.034$ \\
CaII(K)  &393.5 &  $  13.255 \pm  0.525$ & $   1.711 \pm  0.344$ & $   4.026 \pm  0.062$ & $   2.519 \pm  0.499$ & $   6.849 \pm  0.392$ & $   4.698 \pm  0.331$ & $   5.124 \pm  1.317$ & $   7.636 \pm  0.203$ \\
CaII(H)  &397.0 & $  14.027 \pm  0.452$ & $   7.497 \pm  0.678$ & $   4.959 \pm  0.063$ & $   3.971 \pm  1.111$ & $   8.815 \pm  0.357$ & $   8.014 \pm  0.299$ & $   7.775 \pm  1.237$ & $   9.485 \pm  0.201$ \\
CaII  &   850.1 & $ -                 $ & $  29.858 \pm  0.232$ & $ -                 $ & $   4.803 \pm  0.269$ & $ -                 $ & $ -                 $ & $ -                 $ & $ -                 $ \\
CaII  &   866.5 &  $ -                 $ & $  30.195 \pm  0.281$ & $ -                 $ & $   6.345 \pm  0.355$ & $ -                 $ & $ -                 $ & $ -                 $ & $ -                 $ \\
NaI  &    589.2 & $   1.346 \pm  0.067$ & $   1.754 \pm  0.105$ & $   1.148 \pm  0.054$ & $   1.083 \pm  0.211$ & $   1.212 \pm  0.062$ & $   0.344 \pm  0.054$ & $   0.628 \pm  0.109$ & $   2.123 \pm  0.070$ \\
NaI  &    589.8 & $   0.716 \pm  0.068$ & $   0.782 \pm  0.106$ & $   0.729 \pm  0.055$ & $   < 0.534         $ & $   0.693 \pm  0.062$ & $   < 0.144         $ & $   < 0.279         $ & $   0.939 \pm  0.074$ \\
OI  &     777.4 & $   0.759 \pm  0.027$ & $   3.107 \pm  0.128$ & $   0.804 \pm  0.038$ & $   1.013 \pm  0.222$ & $   1.059 \pm  0.049$ & $   0.626 \pm  0.046$ & $   0.736 \pm  0.074$ & $   1.167 \pm  0.055$ \\
OI  &     777.7 & $   0.395 \pm  0.038$ & $   2.955 \pm  0.165$ & $   0.475 \pm  0.036$ & $   1.013 \pm  0.222$ & $   0.680 \pm  0.046$ & $   0.365 \pm  0.042$ & $   0.363 \pm  0.066$ & $   0.645 \pm  0.054$ \\
OI  &     777.8 & $   0.266 \pm  0.042$ & $   2.904 \pm  0.163$ & $   0.273 \pm  0.042$ & $   1.006 \pm  0.221$ & $   0.363 \pm  0.055$ & $   0.153 \pm  0.048$ & $   0.221 \pm  0.073$ & $   0.426 \pm  0.061$ \\
  \enddata
\end{deluxetable*}
\end{rotatetable*}

\noindent performed, and we measured the line flux as we describe in the following.

First of all, while FEROS spectra are Nyquist binned, the ESPRESSO spectra are oversampled. To compute the line flux in a homogeneous way, we rebinned the ESPRESSO spectra in the Nyquist way.
Then, for each line, we fitted the continuum by choosing a suitable local continuum around the line. 
Therefore, we determined the line flux with a python routine which measures the flux of the pixels contained in the area between the fitted continuum and the line. 
We estimated the noise of the line as 
\begin{equation}
    N = \sigma \times \frac{\lambda_{line}}{R} \times \sqrt{N_{\rm pix}}
\end{equation}
where $\sigma$ is the standard deviation of the local continuum, $\lambda_{line}$ is the central wavelength of the line, $R$ is the resolution of the instrument in the corresponding band, and $N_{\rm pix}$ is the number of the pixels contained in the line. 
A line is considered detected when its $S/N > 3$. 
For the non-detected lines, we estimated upper limits for the line flux as: $F^{upp}_{line} = 3 \times N$. 
Results are shown in Tab.\,\ref{tab:flux}.

\subsection{Accretion Luminosity and Mass Accretion Rate} \label{sect:Lacc}

\begin{figure}[t]
    \includegraphics[width=0.9\columnwidth]{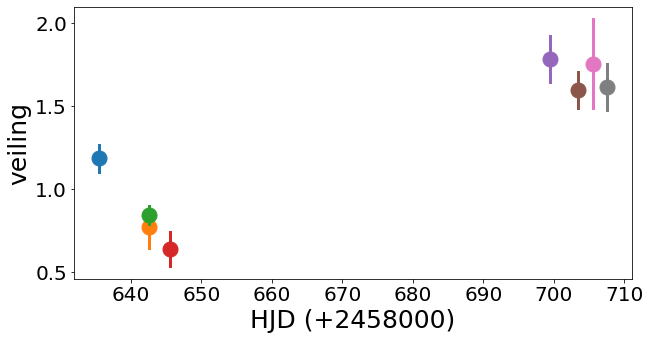}
    \includegraphics[width=\columnwidth]{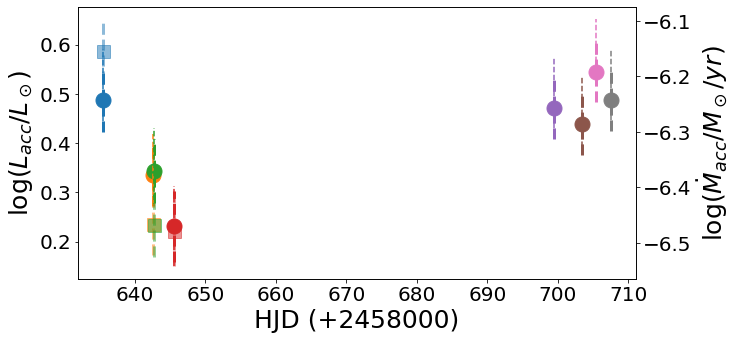}
    \caption{{\it Top}: Absolute veiling measurements for WX\,Cha. 
    {\it Bottom}: The accretion luminosity and the mass accretion rate as a function of the time. The filled circles represent accretion rates computed by using the mean value ($A_V = 3.4 \pm 0.2$). The filled squares represent the accretion rates computed by using the exact value of the extinction for that epoch (see Tab.\,\ref{tab:Av}).}
    \label{fig:veiling}
\end{figure}

We corrected the observed line fluxes of Tab.\,\ref{tab:flux} for the extinction. 
We first used $A_V=3.4$\,mag for all epochs. Then, we took the individual $A_V$ values for when we could estimate them (Epochs\,1, 2, 3 and 4).
In both cases, we converted the resultant fluxes to luminosity, by using the adopted distance. 
Then, we used the empirical relations by \citet{alc17} to compute the  accretion luminosity ($\lacc$) from the line luminosity ($L_{\rm line}$):
\begin{equation}
    \log \big( \lacc/\lsun \big) = a_{\rm line} \log \big( L_{\rm line}/ \lsun \big) + b_{\rm line}
\end{equation}
where $a_{\rm line}$ and $b_{\rm line}$ are coefficients that vary with the line.
We chose the mean value of the accretion luminosities obtained from all the detected lines as the best estimate of the accretion luminosity for WX\,Cha. 
$\lacc$ ranges between $1.58$\,$\lsun$ and $3.16$\,$\lsun$, as shown in Tab.\,\ref{tab:Av} and plotted in Fig.\,\ref{fig:veiling} (bottom panel).

\begin{deluxetable*}{lcccccc}
\tablecaption{\label{tab:Av}}
\tablewidth{0pt}
\tablehead{
\colhead{Epoch} & Veiling & $\log \lacc$ & $\log \macc$ & $A_V$ & $\log \lacc $ & $\log \macc $\\
\colhead{}      &         & $\lsun$ & $\msun$yr$^{-1}$  & mag   & $\lsun$ & $\msun$yr$^{-1}$ \\
\colhead{(1)} & (2) & (3) & (4) & (5) & (6) & (7)}
\startdata
Epoch\,1 & $1.188 \pm 0.081$ & $0.487 \pm 0.064$ & $-6.22 \pm 0.06$ & $3.07^{0.16}_{0.41}$ & $0.588 \pm 0.064$ & $-6.11 \pm 0.06$\\
Epoch\,2 & $0.77 \pm 0.12  $ & $0.335 \pm 0.062$ & $-6.37 \pm 0.06$ & $3.17^{0.17}_{0.46} $ & $0.236 \pm 0.063$ & $-6.47 \pm 0.06$\\
Epoch\,3 & $0.847 \pm 0.054$ & $0.344 \pm 0.065$ & $-6.36 \pm 0.06$ & $3.17^{0.17}_{0.46}$ & $0.233 \pm 0.065$ & $-6.47 \pm 0.07$\\
Epoch\,4 & $0.64 \pm 0.10  $ & $0.232 \pm 0.072$ & $-6.47 \pm 0.07$ & $3.36^{0.17}_{0.46}$ & $0.222 \pm 0.072$ & $-6.48 \pm 0.07$\\
Epoch\,5 & $1.79 \pm 0.14  $ & $0.472 \pm 0.064$ & $-6.23 \pm 0.06$ & - & - & -\\
Epoch\,6 & $1.60 \pm 0.11  $ & $0.439 \pm 0.064$ & $-6.26 \pm 0.06$ & - & - & -\\
Epoch\,7 & $1.75 \pm 0.27  $ & $0.544 \pm 0.060$ & $-6.16 \pm 0.06$ & - & - & -\\
Epoch\,8 & $1.62 \pm 0.14  $ & $0.488 \pm 0.064$ & $-6.21 \pm 0.06$ & - & - & -
\enddata
\tablecomments{ $\lacc$ and $\macc$ in columns (3) and (4) are computed by using $A_V = 3.4 \pm 0.2$\,mag, and the stellar parameters of the primary component computed by \citet[][]{dae13} and scaled at 189.1\,pc. $\lacc$ and $\macc$ in columns (6) and (7) are computed by using the exact value of $A_V$ listed in column (5).}
\end{deluxetable*}

Then, we computed the mass accretion rate $\macc$ using the relation:
\begin{equation}
 \label{eqmacc}
  \macc \sim \left(1 - \frac{\rstar}{R_{\rm in}}\right)^{-1} \frac{\lacc \rstar}{G \mstar}
\end{equation}
where $R_{\rm in}$ is the inner-disk radius which we assume to be $R_{\rm in} \sim 5 R_\star$ \citep{har98}, and $\mstar$ and $\rstar$ are the stellar mass and radius of the primary component \citep[][]{manaraPPVII}. We estimated the absolute error on $\macc$ to be $0.13-0.22$\,dex by propagating the error and taking into account uncertainties on stellar parameters. 
Since we are interested in studying the variability of this system, results listed in Tab.\,\ref{tab:Av} show the error on $\macc$ considering only the uncertainty due to the accretion luminosity. 

The mass accretion rate ranges between $3.31 \times 10^{-7}$\,$\msun/$yr and $7.76 \times 10^{-7}$\,$\msun/$yr during the epochs we observed.
Fig.\,\ref{fig:veiling} (bottom panel) shows the accretion rates of WX\,Cha as a function of the time.
The accretion rate (circles) is large during Epoch\,1, than it decreases, reaching the lowest value during Epoch\,4. 
It increases again in the last four epochs.
This trend is conserved if we study the accretion rates obtained by using different values for the extinction (squares), when it was possible to have a contemporary estimate. In particular, the accretion rate increases during Epoch\,1, dicreases during Epochs\,2 and 3, and remains similar during Epoch\,4. 
Later epochs are uncertain because we do not know the precise $A_V$, even if, for Epoch\,8, the main contribution to the uncertainty is the absence of contemporary photometry, which affects the flux calibration and, therefore, the accretion rate itself. 
However, considering these uncertainties, we can say that the accretion rate increased again during Epoch\,5, decreasing and increasing a little in the following three epochs, ending up similar to Epoch\,1.
We note that the accretion rates computed with the contemporary $A_V$ is in agreement within the error with the ones computed by using the averaged value.
We also note that the variability of the accretion rate follows the same qualitative trend as does the veiling (see both panels of Fig.\,\ref{fig:veiling}).

\section{Discussion} 
\label{sect:discussion}

The main aim of this paper is to investigate for the first time the accretion variability in the WX\,Cha binary system. 
In this section we discuss the light curves and the line profile variations, and we compare our results with other multiple systems and single young stars in the Cha\,I region.

\begin{figure*}[t]
    \centering
    \includegraphics[width=\textwidth]{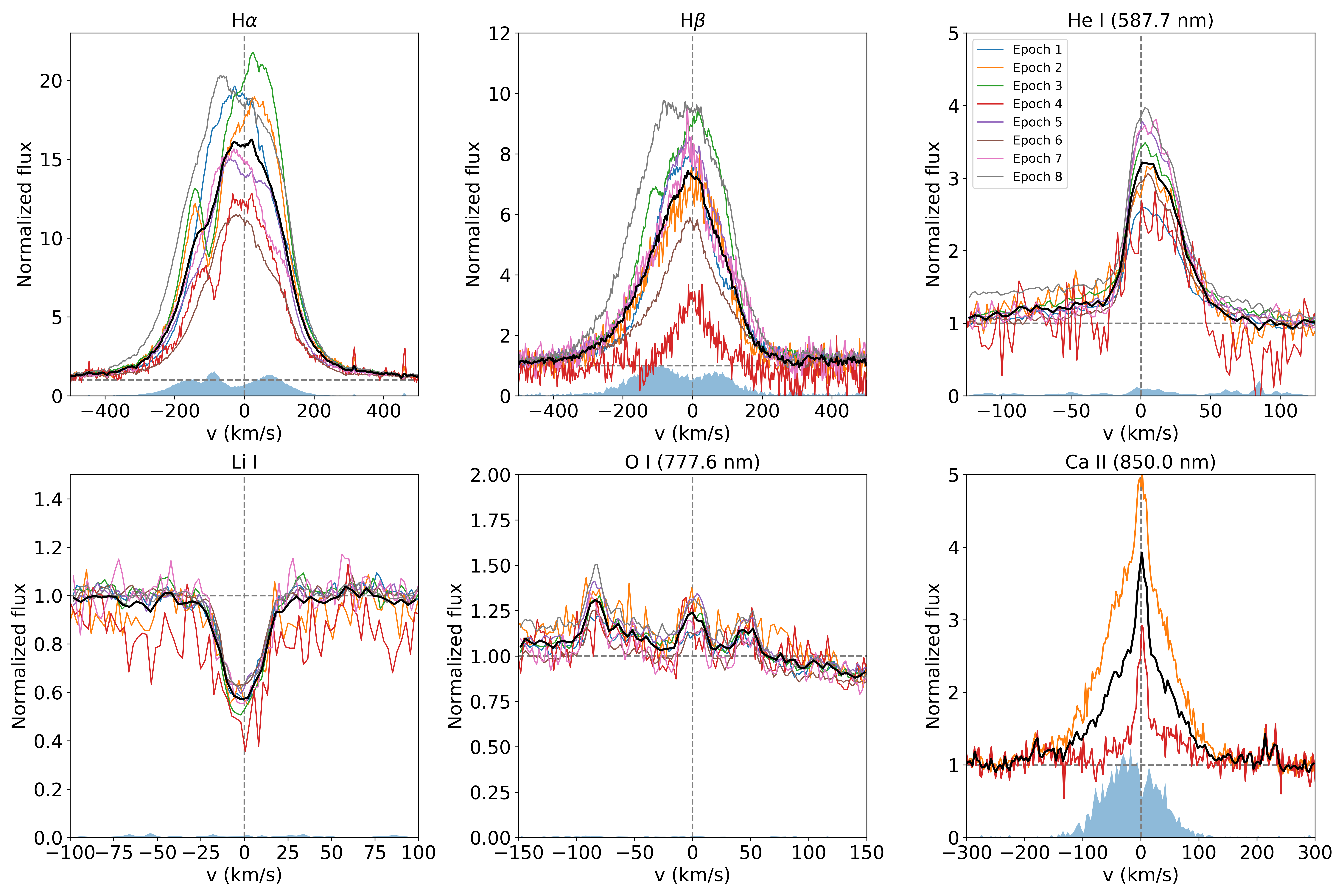}
    \caption{Variance profiles of a few spectral lines. The lines with different colors indicate the different epochs (as listed in the upper right panel), the black thick line shows the average line profile, and the blue shaded area illustrates the variance profile in each panel. The wavelength on the plots are the lines center in the vacuum. \label{fig:var_profile} }
\end{figure*}
\subsection{WX Cha variability}
We described the light curves of WX\,Cha in Sect.\,\ref{sect:period} and the accretion rates in Sect.\,\ref{sect:Lacc}.
Here we investigate the relation between these two quantities.

In Figs.\,\ref{fig:lc_period} and \ref{fig:lc_period_2021} four bright events ($m_{\rm TESS} < 12$\,mag) are present. 
During these events, the variability in the $g$ band reaches 0.8\,mag, while it is only 0.2\,mag in the TESS and NIR light curves.
Such variation in the light curves can be explained by the accretion bursts.
To test this hypothesis, we computed the accretion luminosity and made a quantitative analysis of the accretion contribution during the bright events. 
This was possible for the second brightest event (Epoch\,1), and before (Epoch\,5) and right after the third one (Epoch\,6), when we have spectroscopic data.

Epochs\,1 and 5, and 6 have all large veiling and accretion rates, see Tab.\,\ref{tab:Av}.
The dimmest epochs according to the $g$ band light curve are Epochs\,2, 3 and 4, which have veiling and accretion rates lower than all the other epochs. 
Epoch\,4, in particular, is the one with the lowest accretion rate, and the dimmest in $g$ band. 
This is true both using the fixed value of the extinction and the contemporary one (see Fig.\,\ref{fig:veiling}). 
Since the fluxes during the last four epochs, in particular during the last one, are uncertain due to the less accurate flux calibration (see Sect.\,\ref{sect:spec}), we investigated the variability only for the first four epochs. 
We note that using the contemporary $A_V$ values, when possible, the discrepancy between the light curve and the accretion variability decreases, indeed the central values of Epochs\,2/3 and 4 are more similar if a different value of the extinction is used for the different epochs. 
This hints at the extinction variability.
Indeed the geometrical configuration of the system plays an important role since WX\,Cha has an almost edge-on disk ($i = 87^\circ$). 
This means that depending on the presence and intensity of the accretion funnels, the amount of dust in our line of sight may change significantly on short time scales. 
Moreover, the $M$ coefficent suggests that in 2019 the light curve is symmetric. Since there is accretion variability, as shown in Fig.\,\ref{fig:veiling}, the symmetry of the light curve can be explained by the presence of tiny dips, suggesting extinction variability.
We also looked for other variability sources as tracers typical of ejection flows and winds as H$_2$ and [O{\footnotesize{I}}], but we did not detect them.

\subsection{The variance profile} \label{sect:profile}

Lines profiles contain information about the kinematics of the emitting gas by a certain chemical element. 
Therefore, we plot in Fig.\,\ref{fig:var_profile} a set of emission lines, from top-left to bottom-right: H$\alpha$, H$\beta$, and He{\footnotesize{I}} at $\lambda 587.7$\,nm, Li{\footnotesize{I}} absorption line, O{\footnotesize{I}} triplet, and Ca{\footnotesize{II}}.
All these lines but Li{\footnotesize{I}} are accretion tracers.

The H$\alpha$ line shows what seems a blueshifted bump. 
This can be identified as a low velocity component which peaks around $-190$\,km/s in all the epochs but Epoch\,1 and 8. 
This component is also present in the mean line profile (black curve).
A similar line profile is seen in the spectra of DR\,Tau \citep[][]{alencar2001} and RW\,Aur \citep[][]{alencar2005}.
Even if these sources show similar line profile, their inclinations is different since DR\,Tau is an almost pole-on ($i = 5.4^{2.6}_{2.1}$\,degrees), while the inclination of RW\,Aur is $i = 55.1^{0.5}_{0.4}$ degrees, according to the recent ALMA observations \citep[][]{long2019}.
This profile is due to winds which cause absorption at about $-100$\,km/s.
The presence of outflow is confirmed for both DR\,Tau \citep[][]{facchini2016} and RW\,Aur \citep[][]{giannini2019}, by the presence of winds and jet tracers. On the contrary, we found no such detections in our data. 

The H$\beta$ profile shows the blue low-velocity component only at Epoch\,3. This line profile differs significantly from the H$\beta$ of DR\,Tau and RW\,Aur which show a peaked and spectroscopically separated blue component.
The large noise of FEROS epochs (Epochs\,2 and 4) prevent us from distinguishing the second component in the H$\beta$ line for these epochs.

The He{\footnotesize{I}} at $\lambda 587.7$\,nm shows no variability in the shape, displaying a change only in the strength of the line. 
The line profile of He{\footnotesize{I}} is not symmetric and a red wing up to $40-60$\,km/s is present in every epoch.

The Li{\footnotesize{I}} line and the O{\footnotesize{I}} triplet are almost identical in every epoch. We note that the Li{\footnotesize{I}} line of WX\,Cha, which is similar for all the epochs, is similar in shape to the Li{\footnotesize{I}} line of RW\,Aur and the average line profile (black line in Fig.\,\ref{fig:var_profile}) is as deep as RW\,Aur \citep[][]{facchini2016}.

The wavelength range corresponding to Ca{\footnotesize{II}} triplet was not covered by ESPRESSO, and FEROS covered only the first and the third lines of the triplet, at $\lambda 849.8$\,nm and $\lambda 866.2$\,nm, respectively. Therefore we can study the variability of Ca{\footnotesize{II}} line profile only by comparing our two FEROS spectra, and only for two out of the three lines.
We detected both the two lines during both Epoch\,2 and Epoch\,4. 
We measured their fluxes with the same procedure we used to measure the line flux of the accretion tracers (see Sect.\,\ref{sect:lineflux}) obtaining $F_{Ca{\footnotesize{II}}\,849.8} = (3.184 \pm 0.028) \times 10^{-14}$\,erg\,s$^{-1}$\,cm$^{-2}$ and $F_{Ca{\footnotesize{II}}\,849.8} = (3.11 \pm 0.035) \times 10^{-14}$\,erg\,s$^{-1}$\,cm$^{-2}$ for Epoch\,2; and $F_{Ca{\footnotesize{II}}\,849.8} = (0.515 \pm 0.041) \times 10^{-14}$\,erg\,s$^{-1}$\,cm$^{-2}$ and $F_{Ca{\footnotesize{II}}\,849.8} = (0.562 \pm 0.029) \times 10^{-14}$\,erg\,s$^{-1}$\,cm$^{-2}$ for Epoch\,4. 
The shape of these lines is similar between each other and between the two epochs. 
In Fig.\,\ref{fig:var_profile} we see the Ca{\footnotesize{II}} line profile at $\lambda 849.8$\,nm which is the sum of a narrow and a broad component. Also the shape of this line is similar in DR\,Tau \citep[][]{alencar2001} while in RW\,Aur the line profile is more variable: some epochs show absorption as the H$\alpha$ line, and some epochs show a similar line profile than of WX\,Cha \citep[][]{alencar2005}.
According to \citet[][]{muzerolle.models1998} models, high inclination disks should present prominent redshifted absorption in hydrogen and He lines as a sign of infalling material by the magnetospheric accretion, although it is not a prerequisite and it is highly geometry-dependent. 
Despite the measured inclination disk for WX\,Cha is high \citep[87\,degrees,][]{banzatti2015}, the redshifted absorption is not present in this source lines (see \ref{fig:var_profile}). 
In the \citet[][]{muzerolle.models1998} models the absorption is present with high inclination for $\macc$ up to $10^{-7}$\,$\msun/$yr. 
Since the mass accretion rate we measure is higher than that, assuming the disk inclination estimate is reliable, a possible interpretation of the lack of reshifted absorption features in WX\,Cha can be that this feature may disappear as the accretion rate increases, as it is simply filled up by emission.

The first FEROS spectrum (Epoch\,2) and the second ESPRESSO spectrum (Epoch\,3) were taken at the same night only a few hours apart. 
The two spectra look very similar, as expected, but some strong accretion tracer lines, such as the H$\alpha$ or H$\beta$ lines, show noticeable differences (Fig.\,\ref{fig:var_profile}).
In order to examine whether this discrepancy is an instrumental or a physical effect, we compared several emission and absorption lines in the two spectra. 
We found some differences, however, most lines are in agreement within 10\%. 
Only the strongest lines differ $\sim 30$\%, but as these are typically accretion tracers, the change might have a physical origin.

\subsection{Comparison with Other Chamaeleon Sources}
\label{sect:maccmstar}

\begin{figure}[]
    \centering
    \includegraphics[width=\columnwidth]{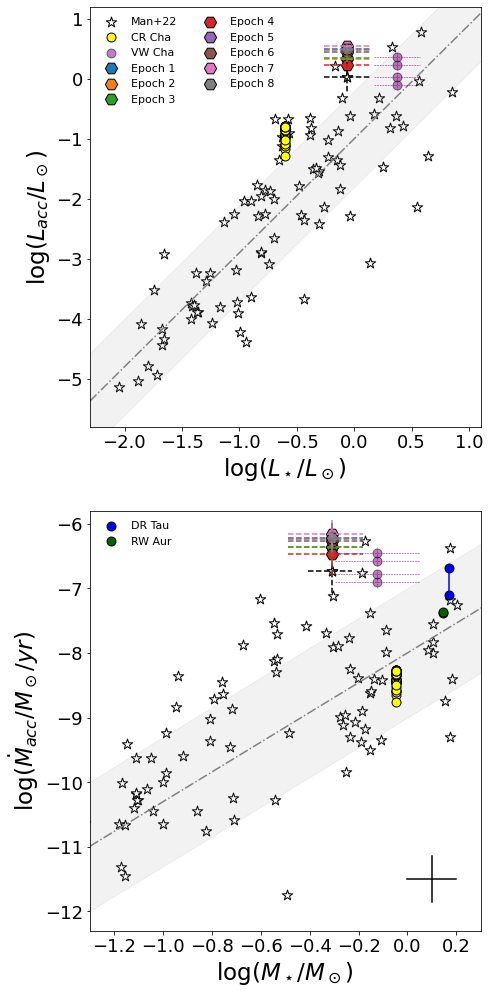}
    \caption{{\it Top}: Accretion luminosity as a function of the stellar luminosity. {\it Bottom}: Mass accretion rate as a function of the stellar mass. Black stars show Cha\,I CTTS sample \citep[][]{manaraPPVII}. In particular, the star with error bar corresponds to WX\,Cha. 
    Yellow, purple, blue and dark green circles correspond to CR\,Cha, VW\,Cha, DR\,Tau, and RW\,Tau accretion rates \citep[][respectively]{zsidi2022, zsidi2022vwcha, giannini2022, facchini2016}. WX\,Cha accretion rates computed in this work are shown in hexagons, each epoch is identified by the a different color as labelled in the legend. The black cross on the bottom-right of the plot represent the uncertainty for Cha\,I CTTS sources. The uncertainties smaller than the symbol size are not presented. Dashed-dotted lines and the grey regions represent the best fit of Cha\,I sources and the relative $1\sigma$ dispersion. \label{fig:acc_stellarpar}}
\end{figure}

\begin{figure}
    \centering
    \includegraphics[width=\columnwidth]{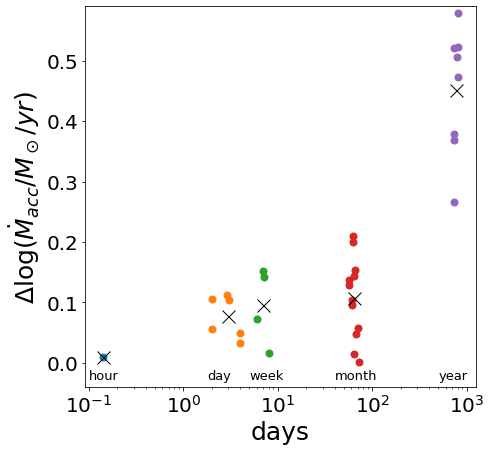}
    \caption{Relative variations in $\macc$, computed from results based on X-Shooter data of this work and \citet[][]{manaraPPVII}, on timescales from hours to 3/4 years. Different colors represent different timescale variations, as described in the legend. Black crosses show the mean value for each timescale sub-sample.}
    \label{fig:deltaMacc}
\end{figure}

To put our results in context, we compare $\lacc$ and $\macc$ with a sample of CTTSs in the Cha\,I star-forming region \citep[empty stars,][]{manaraPPVII}, and with two other sources, part of the same observational effort between TESS photometry and ESPRESSO$+$FEROS spectroscopy, i.e. CR\,Cha and VW\,Cha \citep[][yellow and purple circles, respectively]{zsidi2022, zsidi2022vwcha}.
In Figure\,\ref{fig:acc_stellarpar}, we plot the accretion luminosity as a function of the stellar luminosity in the top panel, and the mass accretion rate as a function of the stellar mass in the bottom panel. The latter plot shows also the $\macc -  \mstar$ relation for DR\,Tau (blue circles) and RW\,Aur (dark green circles) for comparison. Looking at Fig.\,\ref{fig:acc_stellarpar} we see that WX\,Cha accretes more than what is expected based on its luminosity (it is way above the gray stripe in Fig.\,\ref{fig:acc_stellarpar}), and it also accretes more than any of the CTTSs in the Manara et al. sample (stars in Fig.\,\ref{fig:acc_stellarpar}). 

WX\,Cha itself is included in the Manara et al. sample having $\log \lacc = 0.034$ and $\log \macc = -6.729$.  
Both panels of Fig.\,\ref{fig:acc_stellarpar} show that results of WX\,Cha provided by \citet[][]{manaraPPVII} lie below our results on the $\lacc - \lstar$ and $\macc - \mstar$ diagrams, but are compatible with ours within the error. This suggests that a larger variability of the accretion rate is possible on longer time scales than those probed here. 
Indeed, Fig.\,\ref{fig:deltaMacc} shows the variation of the mass accretion time of WX~Cha in different timescales, from hours to years. To develop this plot, we used our data and the results from \citet[][]{manaraPPVII}, whose spectra were observed in 2015 (2015-04-03T07:12:18.1630).
The $\macc$ variation increases from hours to years of about 0.5 dex, higher than the typical variability of 0.3\,dex of the Orion Nebula Cluster \citep[][]{Flaischlen2022}, and smaller than what found for XX\,Cha \citep[2\,dex,][]{Claes2022}. 
Variability timescale from hours to years is in agreement with what found by \citet[][]{costigan12} and \citet[][]{zsidi2022}. 
In contrast, \citet[][]{venuti2015} analyzed the variability timescale of young stars in NGC\,2264 concluding that the typical timescale of variability of this cluster is of the order of days to weeks, not hours to weeks as for WX\,Cha. Therefore timescale variability of WX\,Cha is more rapid that for NGC\,2264 cluster at short timescales.
Unfortunately, we can't test what happens to $\Delta \macc$ for longer timescales, since previous estimate on the mass accretion rate for this source were computed by using different approaches.

Comparing our results with VW\,Cha, we note that this source fits into the general trend outlined by the gray stripe in Fig.\,\ref{fig:acc_stellarpar} (upper panel), while WX\,Cha and CR\,Cha are higher than other CTTS in Cha\,I. 
At the same time, the highest points for VW\,Cha actually have similar $\lacc$ values than the lowest points of WX\,Cha in both the panels.
The $\macc$ of WX\,Cha is in agreement within the error with results of VW\,Cha, and always larger than other Cha\,I CTTSs with the same stellar mass. 
On the contrary, CR\,Cha lies within the trend outlined by the $\macc - \mstar$ Cha\,I distribution.
Among these three sources, only WX\,Cha deviates from the general trend towards both larger accretion luminosity and larger accretion rates. We note that accretion variability alone would not explain the large deviation of WX\,Cha from the general trend since the typical variability for non eruptive CTTS is about 0.5 \citep[][]{lorenzetti2013} which can not justify the fact that $\lacc$ of WX\,Cha is between one and three orders of magnitude higher than for a CTTS in Cha\,I with the same stellar luminosity.
ALMA observations by \citet[][]{pascucci16} reported a flux density $F_\nu =  (20.81 \pm 0.57)$\,mJy in Band\,7 continuum ($887 \mu$m), corresponding to $M_{\rm dust} = 6.02 \, {\rm M}_\Earth$ by assuming a greybody function, the same temperature of 20\,K and a distance of 160\,pc. 
Scaling this value to the distance we adopted, we obtain $M_{\rm dust} = 7.78 \, {\rm M}_\Earth$, which corresponds to what found by \citet[][]{manaraPPVII}. 
Fig.\,6 of \citet[][]{pascucci16} shows that WX\,Cha lies in the portion of the $M_{\rm dust} - \mstar$ distribution that is above the best fit. Therefore, we conclude that WX\,Cha disk mass is large if compared to what is suggested by the best fit for other sources with similar stellar mass. As a consequence, a possible interpretation of the higher mass accretion rate of WX\,Cha and its general deviation from Cha\,I trend can be that this source has more fuel coming from the disk.
It is also possible that the high mass accretion rate for this source is a consequence of the binary nature of WX\,Cha, as found for similar systems by \citet[][]{zagaria2022}.
We also note that the CTTS named 2MASS\,J11095340-7634255,
2MASS\,J11095873-7737088, and
2MASS\,J11040909-7627193 present similar accretion rates and stellar parameters to WX\,Cha system.

\section{Summary}
\label{sect:conclusions}
We performed the first study on the accretion variability of the 0$\farcs$7 separation binary T\,Tauri system WX\,Cha. 
We observed this source with multi-epoch optical and NIR photometry, and high-resolution optical spectroscopy contemporaneously with TESS observations.
According to our analysis, the primary component dominates our measurements, therefore, we studied this binary system as if it was a single source.

The TESS light curve reveals variations on daily timescales with a peak-to-peak amplitude of $\sim$0.5\,mag.
The light curves show four bright events. 
Some of our spectra were taken during these events, and our spectroscopic and photometric analysis together seem to suggest that these events are due to the composite effects of changing accretion rate and variable extinction.

We performed the spectral typing of our system comparing the absorption lines of WX\,Cha to a set of stellar templates, finding that the template with $T_{\rm eff} = 3740$\,K best reproduces our observations. 
This correspond to a M0 spectral type, in agreement with the literature.
From our NIR photometry, we estimated different values of $A_V$. 
We adopted the mean value ($A_V = 3.4 \pm 0.2$\,mag) as the best estimate of the extinction of the system. 

We used the fluxes of different emission lines to measure the accretion rates from our spectra, finding $\log (\lacc / \lsun)$ ranging between 0.2 and 0.5, and $\log (\macc/\msun/{\rm yr})$ between $-6.6$ and $-6.3$. 
Those values are compatible with the most accreting CTTS \citep[$\sim 10^{-12} - 10^{-6.5}\,\msun/$yr, see][]{manaraPPVII}{}, but orders of magnitude larger when compared to the Cha\,I population with similar stellar parameters. We tentatively address this difference to the large disk mass of WX\,Cha, which fuels the accretion process.

The accretion variability follows the same trend as the veiling variability. 
However, the photometric variability cannot be explained by considering only the accretion. Differences between the accretion and light curves variability decrease when we compute the accretion rate using contemporary extinction values. 
Moreover, we measured different $A_V$ values in different epochs. 
This may suggest the presence of $A_V$ variability as well. 
The geometrical configuration of the system with an almost edge-on disk makes this explanation plausible.

We analyzed the profiles and variability of several emission lines in the spectra of WX\,Cha. 
The accretion tracer lines show significant variability both in strength and shape and their profiles are typically complex with multiple components, similarly to the strongly accreting DR\,Tau.

\acknowledgments

We thank the anonymous referee for the careful reading and useful comments they provide, which improve our manuscript.

This project has received funding from the European Research Council (ERC) under the European Union's Horizon 2020 research and innovation programme under grant agreement No 716155 (SACCRED); and under the Marie Sklodowska-Curie grant agreement No 823823 (DUSTBUSTERS).

AB is supported by the Lend\"ulet Program  of the Hungarian Academy of Sciences, project No.  LP2018-7/2021 and the KKP-137523 'SeismoLab' \'Elvonal grant of the Hungarian Research, Development and Innovation Office (NKFIH). Additional support is received from the Hungarian Research, Development and Innovation Office grant K-138962.

This work has made use of data from the European Space Agency (ESA) mission
{\it Gaia} (\url{https://www.cosmos.esa.int/gaia}), processed by the {\it Gaia}
Data Processing and Analysis Consortium (DPAC,
\url{https://www.cosmos.esa.int/web/gaia/dpac/consortium}). Funding for the DPAC
has been provided by national institutions, in particular the institutions
participating in the {\it Gaia} Multilateral Agreement.

This work was partly funded by the Deutsche Forschungsgemeinschaft (DFG, German Research Foundation) - 325594231.

This research received financial support from the project PRIN-INAF 2019 "Spectroscopically Tracing the Disk Dispersal Evolution"

\vspace{5mm}
\facilities{ESO-VLT/XSHOOTER}

\software{astropy \citep{2013A&A...558A..33A},  
          Cloudy \citep{2013RMxAA..49..137F}, 
          SExtractor \citep{1996A&AS..117..393B}
          }

\appendix

\section{SMARTS photometry} \label{app:smarts}

\begin{deluxetable}{lcccccc}
\tablecaption{\label{tab:smart} Near-infrared photometry of WX\,Cha.}
\tablewidth{0pt}
\tablehead{
\colhead{JD} & $V\pm \Delta V$& $R \pm \Delta R$ & $I_C\pm \Delta I_C$ & $J \pm \Delta J$ & $H \pm \Delta H$ & $K \pm \Delta K$ \\
\colhead{$+$2,458,000}    & mag            & mag            & mag             }
\decimalcolnumbers
\startdata
603.066 & $14.293 \pm 0.029 $ & $13.240 \pm 0.002 $ & $12.111 \pm 0.007$ & $9.676 \pm 0.076 $ & $ 8.608  \pm 0.060$ & $ 7.769  \pm 0.048 $ \\
604.066 & $13.928 \pm 0.010 $ & $13.039 \pm 0.009 $ & $11.924 \pm 0.004$ & $9.471 \pm 0.065 $ & $ 8.439  \pm 0.060$ & $ 7.715  \pm 0.073 $ \\
605.078 & $14.152 \pm 0.008 $ & $13.166 \pm 0.009 $ & $12.037 \pm 0.005$ & $9.710 \pm 0.025 $ & $ 8.612  \pm 0.031$ & $ 7.887  \pm 0.061 $ \\
606.039 & $		-			$ & $		-         $	& $12.104 \pm 0.003$ & $9.755 \pm 0.026 $ & $ 8.666  \pm 0.017$ & $ 8.061  \pm 0.108 $ \\
607.027 & $     -           $ & $       -         $ & $12.004 \pm 0.005$ & $9.699 \pm 0.025 $ & $ 8.725  \pm 0.025$ & $ 7.920  \pm 0.062 $ \\
608.051 & $     -           $ & $       -         $ & $11.992 \pm 0.003$ & $9.680 \pm 0.026 $ & $ 8.707  \pm 0.026$ & $ 7.842  \pm 0.050 $ \\
609.039 & $     -           $ & $       -         $ & $11.977 \pm 0.003$ & $9.727 \pm 0.021 $ & $ 8.688  \pm 0.022$ & $ 7.830  \pm 0.048 $ \\
611.031 & $     -           $ & $       -         $ & $12.034 \pm 0.004$ & $9.782 \pm 0.034 $ & $ 8.676  \pm 0.025$ & $ 8.019  \pm 0.044 $ \\
612.051 & $     -           $ & $       -         $ & $12.048 \pm 0.004$ & $9.665 \pm 0.025 $ & $ 8.681  \pm 0.021$ & $ 7.822  \pm 0.035 $ \\
613.020 & $     -           $ & $       -         $ & $12.054 \pm 0.003$ & $9.762 \pm 0.022 $ & $ 8.740  \pm 0.021$ & $ 7.874  \pm 0.031 $ \\
614.055 & $     -           $ & $       -         $ & $12.063 \pm 0.002$ & $9.765 \pm 0.023 $ & $ 8.758  \pm 0.022$ & $ 7.990  \pm 0.037 $ \\
615.031 & $     -           $ & $       -         $ & $11.972 \pm 0.003$ & $9.747 \pm 0.023 $ & $ 8.745  \pm 0.021$ & $ 7.806  \pm 0.046 $ \\
616.031 & $     -           $ & $       -         $ & $12.043 \pm 0.003$ & $9.813 \pm 0.025 $ & $ 8.813  \pm 0.023$ & $ 7.885  \pm 0.052 $ \\
617.059 & $     -           $ & $       -         $ & $11.893 \pm 0.002$ & $9.705 \pm 0.017 $ & $ 8.700  \pm 0.010$ & $ 7.754  \pm 0.160 $ \\
618.055 & $     -           $ & $       -         $ & $11.763 \pm 0.003$ & $9.627 \pm 0.026 $ & $ 8.532  \pm 0.028$ & $ 7.888  \pm 0.048 $ \\
619.023 & $     -           $ & $       -         $ & $11.830 \pm 0.002$ & $9.633 \pm 0.017 $ & $ 8.656  \pm 0.015$ & $ 7.755  \pm 0.033 $ \\
620.020 & $     -           $ & $       -         $ & $11.980 \pm 0.003$ & $9.713 \pm 0.018 $ & $ 8.636  \pm 0.018$ & $ 7.873  \pm 0.028 $ \\
621.043 & $     -           $ & $       -         $ & $12.003 \pm 0.003$ & $9.744 \pm 0.018 $ & $ 8.708  \pm 0.016$ & $ 7.921  \pm 0.026 $ \\
622.016 & $     -           $ & $       -         $ & $12.022 \pm 0.003$ & $9.740 \pm 0.020 $ & $ 8.734  \pm 0.022$ & $ 7.976  \pm 0.028 $ \\
625.027 & $     -           $ & $       -         $ & $12.101 \pm 0.003$ & $9.771 \pm 0.012 $ & $ 8.700  \pm 0.027$ & $ 7.930  \pm 0.070 $ \\
626.027 & $     -           $ & $       -         $ & $12.049 \pm 0.003$ & $9.739 \pm 0.004 $ & $ 8.722  \pm 0.026$ & $ 7.915  \pm 0.066 $ \\
627.031 & $     -           $ & $       -         $ & $11.759 \pm 0.002$ & $9.670 \pm 0.012 $ & $ 8.700  \pm 0.011$ & $ 8.051  \pm 0.032 $ \\
631.082 & $     -           $ & $       -         $ & $11.895 \pm 0.002$ & $9.624 \pm 0.011 $ & $ 8.609  \pm 0.014$ & $ 7.771  \pm 0.104 $ \\
636.039 & $     -           $ & $       -         $ & $11.864 \pm 0.002$ & $9.696 \pm 0.011 $ & $ 8.718  \pm 0.013$ & $ 8.006  \pm 0.029 $ \\
637.020 & $     -           $ & $       -         $ & $11.960 \pm 0.002$ & $9.772 \pm 0.024 $ & $ 8.805  \pm 0.027$ & $ 8.057  \pm 0.013 $ \\
637.992 & $     -           $ & $       -         $ & $12.030 \pm 0.002$ & $9.732 \pm 0.005 $ & $ 8.727  \pm 0.023$ & $ 8.005  \pm 0.085 $ \\
639.047 & $     -           $ & $       -         $ & $12.154 \pm 0.005$ & $9.758 \pm 0.033 $ & $ 8.852  \pm 0.006$ & $ 8.245  \pm 0.064 $ \\
640.023 & $     -           $ & $       -         $ & $12.099 \pm 0.004$ & $9.799 \pm 0.018 $ & $ 8.744  \pm 0.013$ & $ 8.051  \pm 0.093 $ \\
641.004 & $     -           $ & $       -         $ & $12.116 \pm 0.002$ & $9.826 \pm 0.005 $ & $ 8.810  \pm 0.005$ & $ 8.187  \pm 0.054 $ \\
642.023 & $     -           $ & $       -         $ & $12.024 \pm 0.002$ & $9.825 \pm 0.010 $ & $ 8.818  \pm 0.013$ & $ 8.131  \pm 0.036 $ \\
643.000 & $     -           $ & $       -         $ & $12.000 \pm 0.003$ & $9.773 \pm 0.018 $ & $ 8.758  \pm 0.009$ & $ 8.085  \pm 0.063 $ \\
644.016 & $     -           $ & $       -         $ & $12.043 \pm 0.002$ & $9.814 \pm 0.001 $ & $ 8.796  \pm 0.014$ & $ 8.047  \pm 0.056 $ \\
645.000 & $     -           $ & $       -         $ & $12.040 \pm 0.003$ & $9.774 \pm 0.013 $ & $ 8.768  \pm 0.011$ & $ 7.955  \pm 0.049 $ \\\enddata
\end{deluxetable}

In Table\,\ref{tab:smart} we report the SMARTS photometry we observed and used in this work.

\section{Spectra} \label{app:spectra}

\begin{figure*}
    \centering
    \includegraphics[width=1.2\textwidth, angle=90]{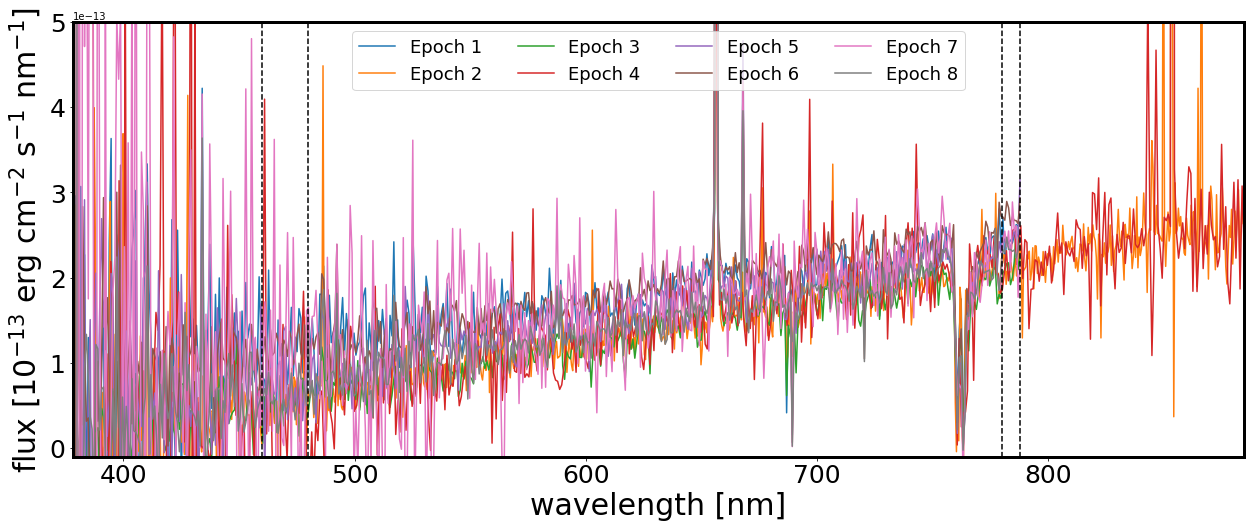}
    \includegraphics[width=1.2\textwidth, angle=90]{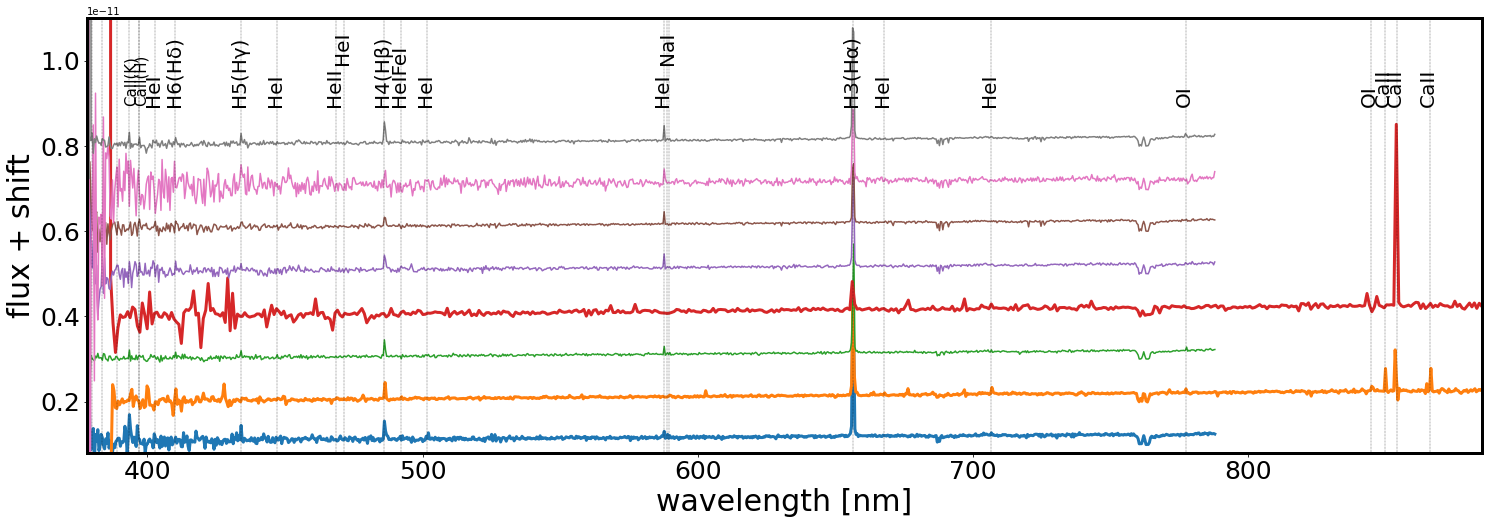}
    \caption{{\it Top} Smoothed flux calibrated spectra of all the epochs.
    {\it Bottom} Smoothed flux calibrated spectra of all the epochs. All the spectra are normalized and suitably shifted in flux for clarity. }
    \label{fig:smooth}
\end{figure*}

Figure\,\ref{fig:smooth} shows the flux calibrated spectra of the 8 epochs we observed.

\bibliography{biblio}{}
\bibliographystyle{aasjournal}

\end{document}